\documentclass[10pt, conference, letterpaper]{IEEEtran}
\usepackage{graphicx,nicefrac} 

\IEEEoverridecommandlockouts
\usepackage{epsfig,rotating,setspace,latexsym,amsmath,epsf,amssymb,amsfonts,bm,bbm,theorem,subfigure,epstopdf,cite,algcompatible,algorithm,caption}

\algrenewcommand\algorithmicforall{\textbf{foreach}}
\algrenewcommand\algorithmicindent{.8em}
 
\usepackage{color,nicefrac}
\usepackage{xcolor}
\usepackage{mathtools}

\newtheorem{theorem}{Theorem}

\newtheorem{problem}{Problem}

\IEEEoverridecommandlockouts
\allowdisplaybreaks
\usepackage{multirow}
\usepackage{soul}

\definecolor{Emerald}{rgb}{0.31, 0.78, 0.47}

\definecolor{RoyalBlue}{cmyk}{1, 0.50, 0, 0}

\definecolor{green1}{rgb}{0.2,0.7,0.2}

\begin{document}

\title{Fully Decentralized Computation Offloading in Priority-Driven Edge Computing Systems\thanks{Emails: \texttt{\{sa57,mazaman2,basar1\}@illinois.edu}, \texttt{bastopcu@}
\texttt{bilkent.edu.tr}, %
\texttt{ulukus@umd.edu}, %
\texttt{akar@ee.bilkent.edu.tr}.
}}

\author{Shubham Aggarwal$^1\!,$ Melih Bastopcu$^2\!,$ Muhammad Aneeq uz Zaman$^1\!,$ Tamer~Ba{\c s}ar$^1\!,$ Sennur Ulukus$^3\!,$ Nail Akar$^2$ \\

$^1$Coordinated Science Laboratory, University of Illinois Urbana Champaign\\

$^2$Department of Electrical and Electronics Engineering, Bilkent University \\

$^3$Department of Electrical and Computer Engineering, University of Maryland}

% \author[1]{Shubham Aggarwal}
% \author[2]{\!Melih Bastopcu}
% \author[1]{\!Muhammad Aneeq uz Zaman}
% \author[1]{\!Tamer Ba{\c s}ar}
% \author[3]{\!Sennur Ulukus}
% \author[2]{\!Nail Akar}

% \affil[1]{Coordinated Science Laboratory, University of Illinois Urbana Champaign}
% \affil[2]{Department of Electrical and Electronics Engineering, Bilkent University}
% \affil[3]{Department of Electrical and Computer Engineering, University of Maryland}
 
\maketitle

\begin{abstract}
We develop a novel framework for fully decentralized offloading policy design in multi access edge computing (MEC) systems. The system comprises $N$ power-constrained user equipments (UEs) assisted by an edge server (ES) to process incoming tasks. Tasks are labeled with urgency flags, and in this paper, we classify them under three urgency levels, namely, high, moderate, and low urgency. We formulate the problem of designing computation decisions for the UEs within a large population noncooperative game framework, where each UE selfishly decides on how to split task execution between its local onboard processor and the ES. We employ the weighted average age of information (AoI) metric to quantify information freshness at the UEs. Increased on-board processing consumes more local power, while increased offloading may potentially incur a higher average AoI due to other UEs' packets being offloaded to the same ES. Thus, we use the mean-field game (MFG) formulation to compute approximate decentralized Nash equilibrium offloading and local computation policies for the UEs to balance between the information freshness and local power consumption. Finally, we provide a projected gradient descent-based algorithm to numerically assess the merits of our approach.   
\end{abstract}

\section{Introduction}
The multi access edge computing (MEC) paradigm has been recently considered as a potential solution to relieve the power-constrained IoT devices of intensive computations. Applications of the MEC system include autonomous driving, medical internet-of-things (IoT), industry 4.0, and large language model (LLM) inference, among others \cite{lin2019computation, feng2022computation, he2024large}. MEC combines technologies from wireless communication and mobile computing to allow resource-limited user equipment (UE) to push compute intensive tasks to the edge of the network. While the traditional cloud computing paradigm can lead to higher transmission/processing delays due to long geographical distances between the cloud and the UEs, the MEC can potentially reduce latency and increase freshness of information due to close proximity to the end-user.

In this work, we propose a fully decentralized offloading policy design approach to improve upon the time responsiveness of the MEC system constituting $N$ UEs and one edge server (ES), and maintain the freshness of information at the UE when \textit{incoming tasks are labeled with urgency flags indicating priority of execution}. Specifically, each incoming task is labeled with a red (R), yellow (Y) or a green (G) flag to denote a high, moderate or low urgency task, respectively. A higher urgency task must meet stricter freshness requirements, and hence, takes priority in service. 
Each UE wishes to make decentralized decisions on the appropriate use of its local processor and the ES facility. We model this computation offloading problem as a noncooperative Nash game (Problem \ref{problem:N_user_game}), for which we provide a low complexity algorithm (Algorithm \ref{alg:MFE}) to compute approximate equilibrium solutions using the mean-field game (MFG) theory \cite{lasry2007mean, huang2007large,al2015joint} (Problem \ref{problem:generic_user}), where the analysis proceeds by considering limiting effects of the UE population. The primary advantage of the MFG setup is that it aids in completely decentralized decision making by using \textit{aggregate population effect} (rather than each individual UE's influence), thereby significantly reducing the overhead of communicating policy information between the ES and UEs.

\textit{Related Work:} Computation offloading in a \textit{single user} mobile-EC system has been studied in \cite{liu2016delay, sathyavageeswaran2024timely} where the policy optimization problem was converted into an optimal programming problem to optimize task latency \cite{liu2016delay} or freshness of information \cite{sathyavageeswaran2024timely}. Within a multi-user scenario, earlier works \cite{mao2016power, li2019computation} adopted a centralized design technique to balance between delay and energy overhead for 3G/4G based wireless systems. Due to high time-complexity of the former, recent approaches have focused on game theoretic frameworks (including both cooperative and noncooperative) \cite{teng2022game, pham2022partial, li2018game, zhou2020partial} for decentralized decision making to achieve selfish and competing objectives of the users. Within this, one line of work considers revenue maximization from the viewpoint of the ESs \cite{teng2022game, pham2022partial}, while another line of work considers equilibrium offloading policy design for the UEs \cite{li2018game, zhou2020partial}. Performance optimization in the latter is then carried out using metrics from queuing theory, such as latency reduction and offloading costs. Alternatively, the works \cite{ma2021freshness, jiang2023age, sathyavageeswaran2024timely} employ the novel metric based on age of information, which balances between the conventional objectives of throughput maximization and delay minimization \cite{kaul2012real}, and maintains freshness of information at the UE. All the above works, however, treat the tasks as a sequence of bits to be forwarded/processed in a sequential manner.

In this work, motivated by resource slicing techniques in 5G and future 6G networks \cite{alsenwi2021intelligent}, we adopt a (large population) game theoretic formulation to model freshness sensitive use cases (such as status updating of environmental information, real-time map navigation and obstacle avoidance in robotic control systems) in \textit{heterogeneous} MEC systems. The heterogeneity within the setup is two-fold: (a) each UE can have different \textit{device parameters}, and (b) each UE receives incoming tasks of \textit{varying urgency levels}, which must then be given different processing priorities. 
To alleviate the problem of computational challenges posed by equilibrium computation in large population games \cite{messous2017computation,li2018game}, and allow for tractable policy design, we appeal to the MFG framework, which has been widely employed in diverse applications such as epidemic control, power grid demand management, financial engineering, wireless networked control systems \cite{olmez2022does, carmona2020applications, aggarwal2024semantic}, to name a few. To the best of authors' knowledge, this is the first work on priority-based MEC systems to compute completely decentralized solutions to the offloading problem without any UE requiring information exchange with other UEs or the ES. 
Finally, we would also like to mention two recent relevant works: \cite{aggarwal2024fully}, which provided a decentralized solution to the offloading problem in MECs with homogeneous tasks; and \cite{akar2024modeling} which provided an alternate absorbing Markov chain-based analysis to compute average AoI expressions for $N$--user shared server systems (such as the MEC systems).

The novelty of our work lies in: (1) A novel theoretically grounded problem formulation for priority-based access, specifically for MEC systems. Such priority driven techniques have been used in other aspects of the wireless systems such as URLLC/eMBB/mMTC resource slicing in 5G new radio (NR) \cite{alsenwi2021intelligent}, but without much theoretical developments. (2) Using the MGF framework to provide completely decentralized tractable and low complexity policy design, which scales very well with a high UE population. Further, our approach provides a clean \textit{recipe} on how to design offloading policies for MEC systems.

\textit{Notation:} $[N]$ denotes the set of UEs $\{1,\ldots,N\}$. Shorthands $\text{Poi}(a)$ and $\text{exp}(a)$ denote Poisson process and exponential random variable, respectively, both with parameter $a$. For a policy vector $a = [a_1, \cdots, a_N]$, $a_{-i}$ denotes the policy vector of all UEs other than $i$.

\section{Problem Formulation}\label{sec:MEC_formulation}
\begin{figure}[t]
    \centering \includegraphics[width=\columnwidth]{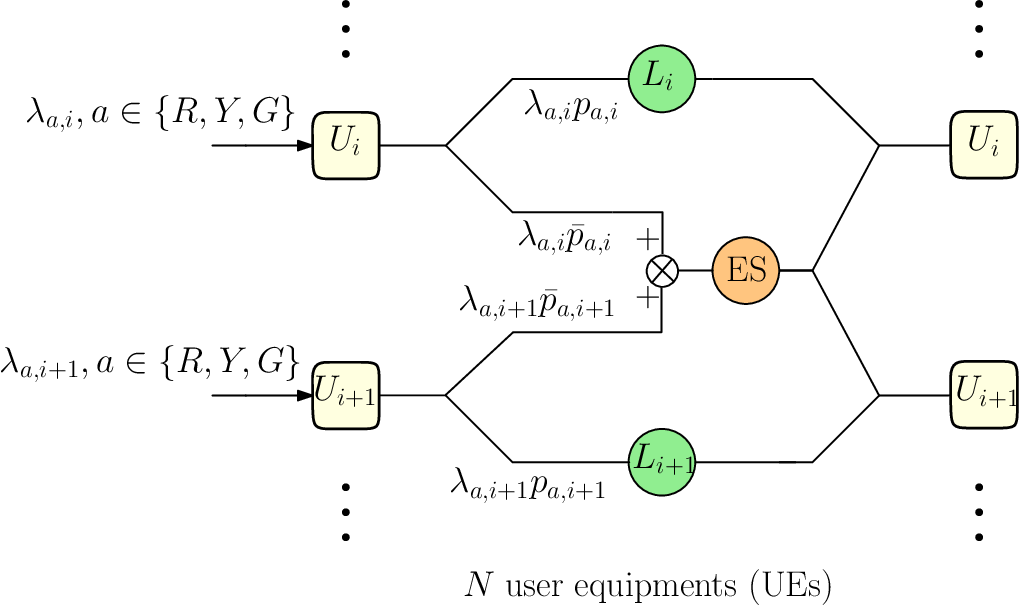}
    \caption{{Flow of incoming tasks for a system of $N$ UEs and shared ES. For UE $U_i$, $L_i$ denotes the UE's local processor.}}
    \label{fig:N_user}
    \vspace{-4mm}
\end{figure}

Consider a MEC system comprising $N$ UEs and an ES as shown in Fig. \ref{fig:N_user}. Each UE $U_i$ can be of a particular \textit{type} $\phi$ which belongs to a finite type set $\Phi$ and has an empirical probability mass function $P_{\mathbb{N}}(\phi)$. The type of a UE allows us to characterize the heterogeneity among UEs, i.e., a UE of type $\phi_1$ can possibly have different parameters such as arrival rates, service rates, etc. compared to a UE of type $\phi_2$. There are tasks arriving at each UE, which need to be processed. Each task is assigned an urgency flag of either \textit{red} (R), \textit{yellow} (Y), or \textit{green} (G). An R flag denotes the highest level of urgency; a Y flag denotes a moderate level of urgency; and a G flag denotes the lowest level of urgency. To execute each task, a UE has two options: (1) serve them onboard using their respective local processors $L_i$, or (2) offload them to the ES, and later download the processed tasks, as shown in Fig. \ref{fig:N_user}. 
We assume that the arrival processes of the different tasks are Poisson with parameters (or intensities) $\lambda_{a,i}$, $a \in \{R,Y,G\}$ for the $i$th UE. The incoming task is then sent to either of the two ($L_i$ or ES) using a Bernoulli random variable with parameters $p_{r,i}$, $p_{y,i}$, and $p_{g,i}$ for red, yellow, and green flagged tasks, respectively. Bernoulli splitting is widely employed in systems with heterogeneous servers \cite{yates2018status} and helps maintain the Poisson property of the tasks for analysis, as we will see in the next sections. We have $0\leq p_{r,i}, p_{y,i}, p_{g,i} \leq 1$. We will use the notation $p_i$ to henceforth denote the triple $[p_{r,i}, p_{y,i}, p_{g,i}]$ for UE $U_i$. Tasks sent to $L_i$ are processed using an $\text{exp}(\mu_{0,i})$ random variable, where $\mu_{0,i}$ denotes the operating frequency of $L_i$ and is bounded as $0 \leq \mu_{0,i} \leq f_{i,max}$, for all $i \in [N]$. Tasks sent to the ES are processed according to an $\text{exp}(\mu)$ random variable, with $\mu \gg \mu_{0,i}, \forall i$.

Henceforth, we will refer to both $L_i$ and ES as servers for all $i$. At each server, we employ the \textit{last-come-first-serve with priority preemption} (LCFS-PP) discipline\footnote{The motivation behind using a preemption based discipline is two-fold: 1) It allows for efficient operation of systems with shared resources and selfish users (quantified using the price of anarchy or the price of stability metrics) as observed in the literature \cite{gai2016packet}, and 2) as we will see later, it allows for a manageable state space for average AoI computation, which can easily increase for a system with a hybrid connection of series-parallel servers.} for task execution, where a task with a higher urgency flag can preempt the one with lower urgency in service, but not vice versa. For instance, a red-flagged task (R-task for short) packet cannot be preempted by any yellow- or green-flagged (Y-/G-task) packets. In addition, an R-task packet from UE $i$ can be preempted by an R-task packet from any UE $j \in [N]$. A similar rule is applied to the Y-/G-tasks as well. The above discipline of priority-based access is motivated by resource slicing techniques in 5G networks \cite{alsenwi2021intelligent}, which constitute packets of URLLC (high priority), mMTC (moderate priority), and eMBB (low priority) type and priority-based transmission protocols are designed to satisfy the delay and accuracy requirements of each. 

When $L_i$ is used by UE $U_i$, onboard local power is consumed according to the relation $P_{\ell,i} = \eta_i \mu_{0,i}^3$, with $\eta_i > 0$ denoting the effective switched capacitance related to the device \cite{mao2016power}. Furthermore, since ES is a shared resource, each UE $j$ serves to decrease freshness of information at UE $i$ since it can preempt some of the packets of UE $i$. For purposes of this work and ease of demonstration, we assume that the ES has only one core/GPU at its disposal, i.e., we do not allow for parallel processing at the ES; such extensions lead to a more cumbersome analysis, left as potential future directions.

Next, we employ the average age of information (AoI) metric to capture the freshness of the information for each UE. Briefly, the AoI at the UE is defined as the time elapsed since the generation of the latest packet received at the receiver, which, in our case, is the UE itself. This metric has been widely used in the literature \cite{kaul2012real, yates2020age, yates2018status} to increase the time responsiveness of the tasks in time-sensitive applications. Thus, each \textit{self-interested} UE is faced with the objective of designing \textit{equilibrium} offloading policies to balance between the local power consumption and maintaining the freshness of information. If a UE decides to operate the local processor at a higher frequency $\mu_{0,i}$, then it can possibly save on the average AoI at the expense of higher power consumption, and vice-versa. Furthermore, as also noted earlier, due to potential packet preemptions at the ES, the AoI of a UE is affected by the offloading policies of the other UEs, and hence the UEs cannot independently optimize their own objectives without appropriately taking into account the \textit{effect} of other UEs in the MEC system. This causes the problem to lie naturally within the class of noncooperative games and is formally introduced next.

Let us define $\boldsymbol{\mu_0}\!:=\! [\mu_{0,1}, \cdots, \mu_{0,N}]$, $\boldsymbol{p_r}\!:=\! [p_{r,1}, \cdots, p_{r,N}]$, $\boldsymbol{p_y}:= [p_{y,1}, \cdots, p_{y,N}]$, $\boldsymbol{p_g}:= [p_{g,1}, \cdots,$ $ p_{g,N}]$, and $\bm{p} = [\bm{p_r}, \bm{p_y}, \bm{p_g}]$.
Then, the multi-objective constrained optimization problem for UE $U_i$ can be formally stated as follows.

\begin{problem}[$N$-user game problem]\label{problem:N_user_game}
    Each UE $U_i$ for $i \in [N]$ aims to minimize its cost $J_i$:
    \begin{align*}
          & \min_{(p_i,\mu_{0,i}) \in [0,1]^3 \times \mathbb{R}} J_i(\boldsymbol{p},\boldsymbol{\mu_0}) := \bar P_{\ell,i} + V \Delta^{(N)}_{i}\!(\boldsymbol{p},\boldsymbol{\mu_0})\\
          & \hspace{1cm} \text{ s.t. } \qquad ~~~~~~ \mu_{0,i} \leq f_{i,max},
    \end{align*}
    with $\Delta^{(N)}_{i}(\boldsymbol{p},\boldsymbol{\mu_0}) :=  w_r \Delta^{(N)}_{r,i}(\boldsymbol{p},\boldsymbol{\mu_0}) + w_y \Delta^{(N)}_{y,i}(\boldsymbol{p},\boldsymbol{\mu_0}) + w_g \Delta^{(N)}_{g,i}(\boldsymbol{p},\boldsymbol{\mu_0})$
    denoting the weighted average AoI incurred by $U_i$ for all task packets, and $\Delta^{(N)}_{r,i}(\boldsymbol{p},\boldsymbol{\mu_0})$, $\Delta^{(N)}_{y,i}(\boldsymbol{p},\boldsymbol{\mu_0})$, and $\Delta^{(N)}_{g,i}(\boldsymbol{p},\boldsymbol{\mu_0})$ denoting the average AoIs of the red, yellow, and the green flagged tasks of UE $i$, respectively, with $w_r, w_y, w_g>0$ being the corresponding relative importance weights. Further, $V>0$ is the scalarization parameter which balances information freshness versus power consumption. A high value of $V$ indicates that the UE cares about time responsiveness more than the power consumed, and vice versa.
    Next, upon performing simple manipulations, the average local power consumed, $\bar P_{\ell,i}$, can be computed as:
    \begin{align*}
        \bar P_{\ell,i} & = [t_{r,i} + (1-t_{r,i})t_{y,i} + (1-t_{r,i})(1-t_{y,i}) t_{g,i}] \eta_i \mu_{0,i}^3, \\
        t_{a,i} & := \frac{\lambda_{a,i} p_{a,i}}{\lambda_{a,i} p_{a,i} + \mu_{0,i}}, ~~ a \in \{R,Y,G\},
    \end{align*}
    where $\!t_{r,i}$, $\!\!t_{y,i}$,$\!$ and $\!\!t_{g,i}\!\!$ are the fraction of time that the local processor is busy servicing R-, Y- and G-task packets, resp.
\end{problem}

The above problem for UE $U_i$ seeks to balance power consumed at $L_i$ versus (relatively weighted) freshness of information, under a hard limit on the operating frequency, by optimizing over the decision variables $p_i$ and $\mu_{0,i}$. To completely formulate the above problem, it only remains to characterize the expression for the average AoI, $\Delta^{N}_i(\bm{p},\bm{\mu_0})$, which we do in the next section. Also, henceforth, we refer to the pair $(p_i,\mu_{0,i})$ as the policy of UE $U_i$.

\section{Average AoI Computation}
We will now compute the expressions for the average AoIs of the red ($\Delta^{(N)}_{r,i}(\boldsymbol{p},\boldsymbol{\mu_0})$), yellow ($\Delta^{(N)}_{y,i}(\boldsymbol{p},\boldsymbol{\mu_0})$), and green ($\Delta^{(N)}_{g,i}(\boldsymbol{p},\boldsymbol{\mu_0})$) tasks. In this regard, we will employ the stochastic hybrid systems (SHS) framework proposed in \cite{hespanha2006modelling,yates2018age}, which we review below very briefly due to page constraints (see \cite{yates2018age} for additional details).

\subsection{Stochastic Hybrid Systems}\label{Subsec:SHS}
\begin{figure}[t]
    \centering \includegraphics[width=0.8\columnwidth]{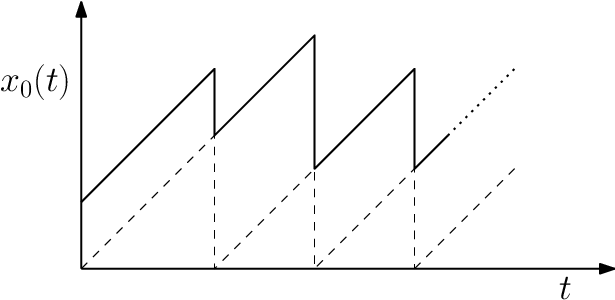}
    \caption{{Evolution of AoI at the UE.}}
    \label{fig:Sampe_AoI}
    \vspace{-3mm}
\end{figure}

Consider a hybrid system whose state $(s(t),x(t)), \forall t \geq 0$ evolves within the space ${\tt S} \times \mathbb R^{n\!+\!1}$ with ${\tt S}$ being a finite set and $n+1$ denoting the number of servers including the UE (which, in our case, equals 3). The partial state $x(t)$ evolves according to a stochastic differential equation in continuous time as
\begin{align*}
    dx(t) = h(t,s,x)dt + k(t,s,x) dB(t),
\end{align*}
where $\{B(t)\}_{t \geq 0}$ is an $(n+1)$-dimensional standard Brownian motion, and $h,k$ are time-varying drift and diffusion coefficients, respectively. Further, the partial state $s(t)$ evolves as a continuous-time (finite) Markov chain (CTMC) from a state $s \in {\tt S}$ to a state $s' \in {\tt S}$ with a transition intensity $q \delta_{s(t^+),s'}$ where $\delta_{a,b} = 1$ if and only if $a = b$, and $t^+$ denotes the infinitesimal instant after time $t$. Further, for each transition of $s(t)$, $x(t)$ also jumps to a value $x'$ according to the rule $x' = \ell(t,s,x)$ at instant $t$. With the above description of a general SHS, the AoI process can be described as its speical (piecewise linear) case with $h(t,s,x) = u_s \in \{0,1\}$, $k(t,s,x) = 0,$ and $\ell(t,s,x) = xA(s)$, where $A(s) \in \{0,1\}^{(n+1) \times (n+1)}$ is a state-dependent constant matrix. A prototypical sample path of the AoI evolution is shown in Fig.~\ref{fig:Sampe_AoI}.

Next, let us define $\pi_{s'}(t):= \mathbb{P}(s(t) = s')$ as the probability of state $s(t) = s'$ and $v_{s'_k}(t):= \mathbb{P}(x_k(t) \mid s(t) = s')$ as the conditional probability of the process $x(t)$ given $s(t) = s'$. Further, let us also define the set of outgoing transitions from a state $s$ as ${\tt O}_s:= \{m: s_{m} = s\}$ and the set of possible incoming transitions to a state $s'$ as ${\tt I}_{s'}:= \{m: s_m = s'\}$. Then, under the assumption of ergodicity of the CTMC, a unique steady state distribution $\Bar{\pi}:= [\bar{\pi}_1, \cdots, \bar{\pi}_c]$ of the CTMC exists \cite{norris1998markov} and satisfies the conservation law:
\begin{subequations}\label{steady_state_prob}
    \begin{align}
    \Bar{\pi}_s \sum_{m \in {\tt O}_s}q^m & = \sum_{m' \in {\tt I}_s} q^{m'} \Bar{\pi}_{s_{m'}}, ~~~~\forall s \in {\tt S}, \\
    \sum_{s \in {\tt S}} \Bar{\pi}_s &= 1, \label{prob_sum}
\end{align}
\end{subequations}
where $c$ denotes the cardinality of ${\tt S}$. Consequently, we invoke the following result from \cite{yates2018age}.
\begin{theorem}\!\cite[Theorem 4]{yates2018age}\label{Avg_Age_thm}
    Suppose $\Bar{\pi}$ is the state distribution of the CTMC and there exists a stationary solution $\Bar{v}\!:=\! [\Bar{v}_1,\! \cdots,\! \Bar{v}_c]$ of the conditional distribution $v_{\cdot}(t)$ satisfying
    \begin{align}\label{cond_prob_eqn}
        \Bar{v}_s \sum_{m \in {\tt O}_s}q^m = u_s \Bar{\pi}_s + \sum_{m' \in {I}_s} q^{m'} \Bar{v}_{s_{m'}}A(m').
    \end{align}
    Then, the average AoI is given by $\Delta = \sum_{s \in {\tt S}}\bar{v}_{s0}$.
\end{theorem}

The SHS review is now thus complete and we will use Theorem~\ref{Avg_Age_thm} to next compute the expressions for the average AoI ${\Delta}^{(N)}_i(\bm{p},\bm{\mu_0})$ of all UE tasks for $i \in [N]$. To do so, for each type of task (R, Y, G), we will first need to construct the CTMC (or equivalently construct a state space and the set of transitions) as described above. Consequently, we will write down the conservation equations \eqref{steady_state_prob} satisfied by the stationary distribution of the constructed CTMC. Finally, we will invoke Theorem \ref{Avg_Age_thm} to state the desired results.

\subsection{Average AoI Calculation for R-Tasks}\label{subsec:red}
Consider the information flow from the perspective of the stream of R-tasks of UE $U_i$ as shown in Fig. \ref{fig:generic_user_red}, where we denote the combined incoming influence of other UEs in Fig. \ref{fig:N_user} using the cumulative rate $\lambda_{-r}$ (as a result of the i.i.d. Bernoulli splitting).  We start by constructing the state space and the corresponding transitions of the CTMC as follows:

\begin{figure}[t]
    \centering
    \includegraphics[width=\columnwidth]{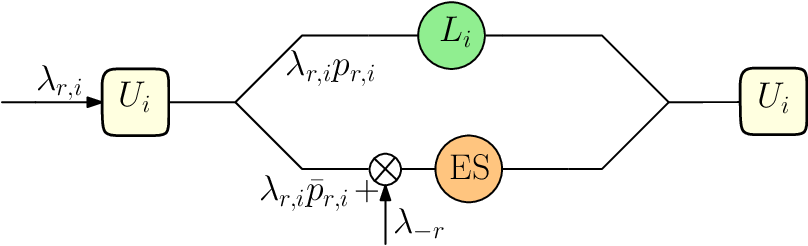}
    \caption{{Task flow for UE $U_i$ from the perspective of R-tasks.}}
    \label{fig:generic_user_red}
    \vspace{-1mm}
\end{figure}

\textit{1) State Space:} First, we make an important observation that the average AoI of the R-tasks is \textit{not} affected by the Y- and G-task packets, which hold a lower priority in service and hence cannot preempt the R-tasks. Thus, from the viewpoint of the R-tasks, it suffices to consider the exogeneous incoming only of the red ones from the other UEs, as shown in Fig. \ref{fig:generic_user_red} with $\lambda_{-r,i}:=\!\! \sum_{j \in [N], j \ne i} \lambda_{r,j} \bar p_{r,j}$. Since we follow a preemption type of service discipline, we use the fake update approach \cite{yates2018status} in both the servers ($L_i$ and ES) to keep the state space manageable. This means that even if a server becomes empty when a packet departs after finishing service, for purposes of AoI computation, we can ignore its ``idleness'' and instead  start servicing a fake packet with its age the same as that of the departing packet. This then allows us to completely describe the state space of the corresponding CTMC using 3 states as detailed in Table \ref{table:states_red}, which track the server servicing the freshest task and a task of other UE (which we refer to as ``class 2'' task). Consequently, following the discussion of the previous subsection, we also have that $u_s = [1~1~1], \forall s \in {\tt S}$.

\textit{2) State Transitions:} The state transitions corresponding to the state space described above are listed in Table \ref{table:transitions_red}, where we keep simultaneous track of the AoIs of the packets at the UE (denoted as $x_0$), and in $L_i$ (denoted as $x_1$) and ES (denoted as $x_2$) using the vector $x(t):= [x_0(t),x_1(t),x_2(t)]$. The value $u_s = [1~1~1]$ then denotes that all three servers (UE, $L_i$, ES) are always busy with a real or a fake packet. We may henceforth forego the use of subscript $i$ if clear from context.

\begin{table}[h!]
    \centering
    \begin{tabular}{ |c|c|c|c| } 
    \hline
    state & server 1 ($L_i$) & server 2 (ES) \\
    \hline
    $s_1$ & freshest  & $2^{nd}$ freshest \\ \hline
    $s_2$ & $2^{nd}$ freshest & freshest \\ \hline
    $s_3$ & freshest & class 2 \\ \hline
    \end{tabular}
    \caption{{The state dictionary for the finite CTMC.}}
    \label{table:states_red}
\end{table}

\begin{table}[h]
    \centering
    %\small
    \begin{tabular}{|c|c|c|l|l|} 
    \hline
    $s$ & $q$ & $s'$ & $x' = xA(s)$ & $v_sA(s)$\\
    \hline
    \multirow{4}{2em}{$s_1$} & $\lambda_r p_r$ & $s_1$ & $[x_0 ~0 ~ x_2]$ & $[\bar{v}_{10}~0~\bar{v}_{12}]$\\
    & $\lambda_r\Bar{p}_r$ & $s_2$ & $[x_0 ~x_1 ~0]$ & $[\bar{v}_{10}~\bar{v}_{11}~0]$ \\
    & $\lambda_{-r}$ & $s_3$ & $[x_0 ~x_1 ~x_0]$ & $[\bar{v}_{10}~\bar{v}_{11}~\bar{v}_{10}]$ \\
    & $\mu_0$ & $s_1$ & $[x_1 ~x_1 ~x_1]$ & $[\bar{v}_{11}~\bar{v}_{11}~\bar{v}_{11}]$ \\
    & $\mu$ & $s_1$ & $[x_2 ~x_1 ~x_2]$ & $[\bar{v}_{12}~\bar{v}_{11}~\bar{v}_{12}]$ \\ \hline
    \multirow{4}{2em}{$s_2$} & $\lambda_r p_r$ & $s_1$ & $[x_0 ~0 ~ x_2]$ & $[\bar{v}_{20}~0~\bar{v}_{22}]$\\
    & $\lambda_r\Bar{p}_r$ & $s_2$ & $[x_0 ~x_1 ~0]$ & $[\bar{v}_{20}~\bar{v}_{21}~0]$ \\
    & $\lambda_{-r}$ & $s_3$ & $[x_0 ~x_1 ~x_0]$ & $[\bar{v}_{20}~\bar{v}_{21}~\bar{v}_{20}]$ \\
    & $\mu_0$ & $s_2$ & $[x_1 ~x_1 ~x_2]$ & $[\bar{v}_{21}~\bar{v}_{21}~\bar{v}_{22}]$ \\
    & $\mu$ & $s_2$ & $[x_2 ~x_2 ~x_2]$ & $[\bar{v}_{22}~\bar{v}_{22}~\bar{v}_{22}]$ \\ \hline
    \multirow{4}{2em}{$s_3$} & $\lambda_r p_r$ & $s_3$ & $[x_0 ~0 ~x_2]$ & $[\bar{v}_{30}~0~\bar{v}_{32}]$\\
    & $\lambda_r\Bar{p}_r$ & $s_2$ & $[x_0 ~x_1 ~0]$ & $[\bar{v}_{30}~\bar{v}_{31}~0]$ \\
    & $\lambda_{-r}$ & $s_3$ & $[x_0 ~x_1 ~x_0]$ & $[\bar{v}_{30}~\bar{v}_{31}~\bar{v}_{30}]$ \\
    & $\mu_0$ & $s_3$ & $[x_1 ~x_1 ~x_1]$ & $[\bar{v}_{31}~\bar{v}_{31}~\bar{v}_{31}]$ \\
    & $\mu$ & $s_3$ & $[x_2 ~x_1 ~x_2]$ & $[\bar{v}_{32}~\bar{v}_{31}~\bar{v}_{32}]$ \\ \hline
    \end{tabular}
    \caption{{The CTMC state transitions and associated AoI jumps.}}
    \label{table:transitions_red}
    \vspace{-2mm}    
\end{table}

Now that the CTMC is completely characterized by Tables \ref{table:states_red} and \ref{table:transitions_red}, we proceed toward computing the average AoI for UE $U_{i}$. In this regard, let us define the quantity $\varrho_r:= \lambda_r + \lambda_{-r} + \mu_0 + \mu$. Then, using \eqref{steady_state_prob}, the steady state vector for R-tasks $\Bar{\pi}_r$ satisfies \eqref{prob_sum} and the following set of equations:
\begin{subequations}\label{steady_state_prob_red}
    \begin{align}
        \varrho_r \Bar{\pi}_{r,1} & = (\lambda_r p_r + \mu_0 + \mu)\Bar{\pi}_{r,1} + \lambda_r p_r \Bar{\pi}_{r,2}, \\ 
        \varrho_r \Bar{\pi}_{r,2} & = (\lambda_r\Bar{p}_r + \mu_0 + \mu)\Bar{\pi}_{r,2} + \lambda_r\Bar{p}_r (\bar \pi_{r,1} + \Bar{\pi}_{r,3}), \\ 
        \varrho_r \Bar{\pi}_{r,3} & = (\lambda_r p_r \!+\! \lambda_{-r} \!+\! \mu_0 \!+\! \mu) \Bar{\pi}_{r,3} + \lambda_{-r} (\Bar{\pi}_{r,1} + \Bar{\pi}_{r,2}).
    \end{align}
\end{subequations}
    
Consequently, using \eqref{cond_prob_eqn}, we can write the set of linear equations satisfied by the steady-state conditional distribution vector as in \eqref{eqn_v_s_red} below: 
\begin{align}\label{eqn_v_s_red}
    \varrho_r \Bar{v}_1 \! \!=& u_s \Bar{\pi}_{r,1} \!+\! \lambda_r{p}_r [\bar{v}_{10}~0~\bar{v}_{12}] + \mu_0 [\bar{v}_{11}~\bar{v}_{11}~\bar{v}_{11}] \nonumber \\
    & + \mu [\bar{v}_{12}~\bar{v}_{11}~\bar{v}_{12}] + \lambda_r p_r [\bar{v}_{20}~0~\bar{v}_{22}],\nonumber \\
    \varrho_r \Bar{v}_2 \!\! =& u_s \Bar{\pi}_{r,2} + \lambda_r \Bar{p}_r [\bar{v}_{20}~\bar{v}_{21}~0] + \mu_0 [\bar{v}_{21}~\bar{v}_{21}~\bar{v}_{22}] \\
    & + \mu [\bar{v}_{22}~\bar{v}_{22}~\bar{v}_{22}] \!+\! \lambda_r \Bar{p}_r [\bar{v}_{10}~\bar{v}_{11}~0] \!+\! \lambda_r \Bar{p}_r [\bar{v}_{30}~\bar{v}_{31}~0], \nonumber \\
    \!\varrho_r \Bar{v}_3  \!\!= &  u_s \Bar{\pi}_{r,3} + \lambda_r p_r [\bar{v}_{30}~ 0 ~\bar{v}_{32}] \!+\! \lambda_{-r} [\bar{v}_{30}~ \bar{v}_{31}~ \bar{v}_{30}] \nonumber \\
    & +\!\mu [\bar{v}_{32}~\bar{v}_{31}~\bar{v}_{32}] \!\!+\!\! \lambda_{-r}[\bar{v}_{10}~\bar{v}_{11}~\bar{v}_{10}]  \!\!+\!\! \lambda_{-r}[\bar{v}_{20}~\bar{v}_{21}~\bar{v}_{20}] \nonumber \\
    & + \!\mu_0 [\bar{v}_{31}~ \bar{v}_{31} ~\bar{v}_{31}]. \nonumber
\end{align}
    
Next, we state the following result providing the expression of average AoI for the R-tasks ($\Delta^{(N)}_{r,i}(\bm{p},\bm{\mu_0})$) for UE $U_i$.

\begin{theorem}\label{thm:AoI_MEC_red}
    Suppose that the arrival rate of red tasks at UE $U_i$ is distributed as $\text{Poi}(\lambda_{r,i})$ and the service rates of $L_i$ and ES are distributed as $\text{exp}(\mu_{0,i})$ and $\text{exp}(\mu)$, respectively. Then, the average AoI ${\Delta}^{(N)}_{r,i}(\bm{p},\bm{\mu_0})$ is given in \eqref{AoI_red_task}.
\end{theorem}

\begin{figure*}[h]
    \begin{align}\label{AoI_red_task}
        \Delta^{(N)}_{r,i}(\bm{p},\bm{\mu_0}) &  = \frac{\mu_{0,i} (\lambda_{-r} + \mu)(\lambda_{-r} + \mu + \mu_{0,i}) + \lambda_{r,i}^3 p_{r,i} \bar p_{r,i} + \lambda_{r,i}^2(\mu + \mu_{0,i}+\lambda_{-r}p_{r,i}(2-p_{r,i}))}{\lambda_{r,i}(\mu + (\lambda_{r,i} + \lambda_{-r})p_{r,i})(\mu_{0,i}(\lambda_{-r} + \mu + \mu_{0,i}) + \lambda_{r,i}(\mu + \mu_{0,i})\bar p_{r,i} )} \nonumber \\
        & \hspace{3.5cm} + \frac{\lambda_{r,i}((\mu + \mu_{0,i})^2+ \lambda_{-r}^2 p_{r,i} + \lambda_{-r}(\mu(1+p) + 2\mu_{0,i})}{\lambda_{r,i}(\mu + (\lambda_{r,i} + \lambda_{-r})p_{r,i})(\mu_{0,i}(\lambda_{-r} + \mu + \mu_{0,i}) + \lambda_{r,i}(\mu + \mu_{0,i})\bar p_{r,i} )}.
    \end{align}
    \hrule
\end{figure*}

The proof of Theorem~\ref{thm:AoI_MEC_red} follows by explicitly solving the set of linear equations \eqref{steady_state_prob_red} to get $\bar{\pi}_{r,i}$, substituting them in the set of equations given by \eqref{eqn_v_s_red} and solving the latter. 

The computation of ${\Delta}^{(N)}_{r,i}(\bm{p},\bm{\mu_0})$
is now complete and we next compute the average AoI for the Y- and the G-tasks.

\subsection{Average AoI Calculation for Y-Tasks and G-Tasks}\label{subsec:yellow}
\begin{figure}[t]
    \centering
    \includegraphics[width=\columnwidth]{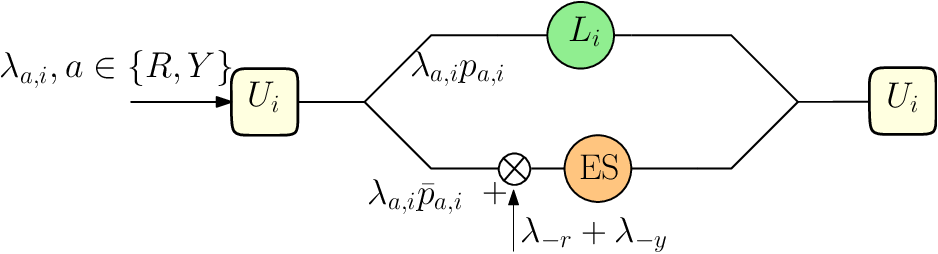}
    \caption{{Task flow for UE $U_i$ from the perspective of Y-/G-tasks.}}
    \label{fig:generic_user_yellow}
    \vspace{-3mm}
\end{figure}

Here, we will provide a common analysis for the yellow and green tasks. To this end, similar to the case with R-tasks, let us consider the information flow from the perspective of either Y-tasks or G-tasks of UE $U_i$ as shown in Fig. \ref{fig:generic_user_yellow}. In the figure, we define the exogenous incoming tasks of same urgency level as $\lambda_e$. In particular, if one is looking at Y-tasks, we have $\lambda_e = \lambda_{-y} = \sum_{j \in [N], j \ne i} \lambda_{y,j} \bar p_{y,j}$. The higher urgency exogenous tasks are handled later. The lower urgency green ones can be ignored since they would not affect the AoI of the Y-tasks. On the other hand, if one is looking from the perspective of G-tasks, we have $\lambda_e = \lambda_{-g} = \sum_{j \in [N], j \ne i} \lambda_{g,j} \bar p_{g,j}$. Then, we can construct the state space and the state transitions of the corresponding CTMC as follows.

\textit{1) State Space:} Similar to the case with R-tasks as detailed earlier, we run fake packets of the lowest priority in both the servers ($L_i$ and ES) with the age of the fake packet equal to the age of the departing packet. Consequently, we describe the state space of the corresponding CTMC (for both the Y and G-tasks) using 7 states as detailed in Table \ref{table:states_yellow}, which track the server servicing the freshest (Y/G) task of the UE $U_i$, a higher priority task (which we refer to as ``class h'' task and constitutes R-tasks when considering the case of Y-tasks, and R- and Y-tasks when considering the case of G-tasks) and same priority task of other UEs (which we refer to as ``class 2'' task). Consequently, we also have that $u_s = [1~1~1], \forall s \in {\tt S}$.

\textit{2) State Transitions:} The state transitions corresponding to the state space described above are listed in Table \ref{table:transitions_yellow}, where we also simultaneously track the AoIs of the packets at the UE ($x_0$), and in $L_i$ ($x_1$) and ES ($x_2$), using the vector $x(t)$. Within this, we use the notation $\lambda^{b,h}, b \in \{Y,G\}$ to denote the total incoming rate of tasks with higher priority, i.e., for the Y-tasks,  $\lambda^{y,h}:= \sum_{j \in [N]} \lambda_{r,j} \bar p_{r,j}$ while for the G-tasks, $\lambda^{g,h}:= \sum_{j \in [N]} \lambda_{r,j} \bar p_{r,j} + \lambda_{y,j} \bar p_{y,j}$. Furthermore, we also use $\lambda^L_{b,h}$ to denote the  rate of tasks of a higher priority which arrive at the local processor. In particular, we have that $\lambda^L_{y,h} = \lambda_r p_r$ for Y-tasks and $\lambda^L_{g,h}=\lambda_r p_r + \lambda_y p_y$ for G-tasks. 

\begin{table}[h]
    \centering
    \begin{tabular}{ |c|c|c|c| } 
    \hline
    state & server 1 ($L_i$) & server 2 (ES) \\
    \hline
    $s_1$ & freshest  & $2^{nd}$ freshest \\ \hline
    $s_2$ & $2^{nd}$ freshest & freshest \\ \hline
    $s_3$ & freshest & class h \\ \hline
    $s_4$ & freshest & class 2 \\ \hline
    $s_5$ & class h & freshest \\ \hline
    $s_6$ & class h & class h \\ \hline
    $s_7$ & class h & class 2 \\ \hline
    \end{tabular}
    \caption{{State dictionary for the finite CTMC for Y/G-tasks.}}
    \label{table:states_yellow}
    % \vspace{-4mm}
\end{table}

\begin{table}[h]
    \centering
    \small
    \begin{tabular}{|c|c|c|l|l|} 
    \hline
    $s$ & $q$ & $s'$ & $x' = xA(s)$ & $v_sA(s)$\\
    \hline
    \multirow{4}{2em}{$s_1$} & $\lambda_b p_b$ & $s_1$ & $[x_0 ~0 ~ x_2]$ & $[\bar{v}_{10}~0~\bar{v}_{12}]$\\
    & $\lambda_b \Bar{p}_b$ & $s_2$ & $[x_0 ~x_1 ~0]$ & $[\bar{v}_{10}~\bar{v}_{11}~0]$ \\
    & $\lambda^L_{b,h}$ & $s_5$ & $[x_0 ~x_0 ~x_2]$ & $[\bar{v}_{10}~\bar{v}_{10}~\bar{v}_{12}]$ \\
    & $\lambda^{b,h}$ & $s_3$ & $[x_0 ~x_1 ~x_0]$ & $[\bar{v}_{10}~\bar{v}_{11}~\bar{v}_{10}]$ \\
    & $\lambda_{e}$ & $s_4$ & $[x_0 ~x_1 ~x_0]$ & $[\bar{v}_{10}~\bar{v}_{11}~\bar{v}_{10}]$ \\
    & $\mu_0$ & $s_1$ & $[x_1 ~x_1 ~x_1]$ & $[\bar{v}_{11}~\bar{v}_{11}~\bar{v}_{11}]$ \\
    & $\mu$ & $s_1$ & $[x_2 ~x_1 ~x_2]$ & $[\bar{v}_{12}~\bar{v}_{11}~\bar{v}_{12}]$ \\ \hline
    \multirow{4}{2em}{$s_2$} & $\lambda_b p_b$ & $s_1$ & $[x_0 ~0 ~ x_2]$ & $[\bar{v}_{20}~0~\bar{v}_{22}]$\\
    & $\lambda_b\Bar{p}_b$ & $s_2$ & $[x_0 ~x_1 ~0]$ & $[\bar{v}_{20}~\bar{v}_{21}~0]$ \\
    & $\lambda^L_{b,h}$ & $s_5$ & $[x_0 ~x_0 ~x_2]$ & $[\bar{v}_{20}~\bar{v}_{20}~\bar{v}_{22}]$ \\
    & $\lambda^{b,h}$ & $s_3$ & $[x_0 ~x_1 ~x_0]$ & $[\bar{v}_{20}~\bar{v}_{21}~\bar{v}_{20}]$ \\
    & $\lambda_{e}$ & $s_4$ & $[x_0 ~x_1 ~x_0]$ & $[\bar{v}_{20}~\bar{v}_{21}~\bar{v}_{20}]$ \\
    & $\mu_0$ & $s_2$ & $[x_1 ~x_1 ~x_2]$ & $[\bar{v}_{21}~\bar{v}_{21}~\bar{v}_{22}]$ \\
    & $\mu$ & $s_2$ & $[x_2 ~x_2 ~x_2]$ & $[\bar{v}_{22}~\bar{v}_{22}~\bar{v}_{22}]$ \\ \hline
    \multirow{4}{2em}{$s_3$} & $\lambda_b p_b$ & $s_3$ & $[x_0 ~0 ~ x_2]$ & $[\bar{v}_{30}~0~\bar{v}_{32}]$\\
    & $\lambda_b\Bar{p}_b$ & $s_3$ & $[x_0 ~x_1 ~x_2]$ & $[\bar{v}_{30}~\bar{v}_{31}~\bar{v}_{32}]$ \\
    & $\lambda^L_{b,h}$ & $s_6$ & $[x_0 ~x_0 ~x_2]$ & $[\bar{v}_{30}~\bar{v}_{30}~\bar{v}_{32}]$ \\
    & $\lambda^{b,h}$ & $s_3$ & $[x_0 ~x_1 ~x_0]$ & $[\bar{v}_{30}~\bar{v}_{31}~\bar{v}_{30}]$ \\
    & $\lambda_{e}$ & $s_3$ & $[x_0 ~x_1 ~x_2]$ & $[\bar{v}_{30}~\bar{v}_{31}~\bar{v}_{32}]$ \\
    & $\mu_0$ & $s_3$ & $[x_1 ~x_1 ~x_1]$ & $[\bar{v}_{31}~\bar{v}_{31}~\bar{v}_{31}]$ \\
    & $\mu$ & $s_4$ & $[x_2 ~x_1 ~x_2]$ & $[\bar{v}_{32}~\bar{v}_{31}~\bar{v}_{32}]$ \\ \hline
    \multirow{4}{2em}{$s_4$} & $\lambda_b p_b$ & $s_4$ & $[x_0 ~0 ~ x_2]$ & $[\bar{v}_{40}~0~\bar{v}_{42}]$\\
    & $\lambda_b\Bar{p}_b$ & $s_2$ & $[x_0 ~x_1 ~0]$ & $[\bar{v}_{40}~\bar{v}_{41}~0]$ \\
    & $\lambda^L_{b,h}$ & $s_7$ & $[x_0 ~x_0 ~x_2]$ & $[\bar{v}_{40}~\bar{v}_{40}~\bar{v}_{42}]$ \\
    & $\lambda^{b,h}$ & $s_3$ & $[x_0 ~x_1 ~x_0]$ & $[\bar{v}_{40}~\bar{v}_{41}~\bar{v}_{40}]$ \\
    & $\lambda_{e}$ & $s_4$ & $[x_0 ~x_1 ~x_0]$ & $[\bar{v}_{40}~\bar{v}_{41}~\bar{v}_{40}]$ \\
    & $\mu_0$ & $s_4$ & $[x_1 ~x_1 ~x_1]$ & $[\bar{v}_{41}~\bar{v}_{41}~\bar{v}_{41}]$ \\
    & $\mu$ & $s_4$ & $[x_2 ~x_1 ~x_2]$ & $[\bar{v}_{42}~\bar{v}_{41}~\bar{v}_{42}]$ \\ \hline
    \multirow{4}{2em}{$s_5$} & $\lambda_b p_b$ & $s_5$ & $[x_0 ~x_1 ~ x_2]$ & $[\bar{v}_{50}~\bar{v}_{51}~\bar{v}_{52}]$\\
    & $\lambda_b\Bar{p}_b$ & $s_5$ & $[x_0 ~x_1 ~0]$ & $[\bar{v}_{50}~\bar{v}_{51}~0]$ \\
    & $\lambda^L_{b,h}$ & $s_5$ & $[x_0 ~x_0 ~x_2]$ & $[\bar{v}_{50}~\bar{v}_{50}~\bar{v}_{52}]$ \\
    & $\lambda^{b,h}$ & $s_6$ & $[x_0 ~x_1 ~x_0]$ & $[\bar{v}_{50}~\bar{v}_{51}~\bar{v}_{50}]$ \\
    & $\lambda_{e}$ & $s_7$ & $[x_0 ~x_1 ~x_0]$ & $[\bar{v}_{50}~\bar{v}_{51}~\bar{v}_{50}]$ \\
    & $\mu_0$ & $s_2$ & $[x_1 ~x_1 ~x_2]$ & $[\bar{v}_{51}~\bar{v}_{51}~\bar{v}_{52}]$ \\
    & $\mu$ & $s_5$ & $[x_2 ~x_2 ~x_2]$ & $[\bar{v}_{52}~\bar{v}_{52}~\bar{v}_{52}]$ \\ \hline
    \multirow{4}{2em}{$s_6$} & $\lambda_b p_b$ & $s_6$ & $[x_0 ~x_1 ~ x_2]$ & $[\bar{v}_{60}~\bar{v}_{61}~\bar{v}_{62}]$\\
    & $\lambda_b\Bar{p}_b$ & $s_6$ & $[x_0 ~x_1 ~ x_2]$ & $[\bar{v}_{60}~\bar{v}_{61}~\bar{v}_{62}]$ \\
    & $\lambda^L_{b,h}$ & $s_6$ & $[x_0 ~x_0 ~x_2]$ & $[\bar{v}_{60}~\bar{v}_{60}~\bar{v}_{62}]$ \\
    & $\lambda^{b,h}$ & $s_6$ & $[x_0 ~x_1 ~x_0]$ & $[\bar{v}_{60}~\bar{v}_{61}~\bar{v}_{60}]$ \\
    & $\lambda_{e}$ & $s_6$ & $[x_0 ~x_1 ~x_2]$ & $[\bar{v}_{60}~\bar{v}_{61}~\bar{v}_{62}]$ \\
    & $\mu_0$ & $s_3$ & $[x_1 ~x_1 ~x_2]$ & $[\bar{v}_{61}~\bar{v}_{61}~\bar{v}_{62}]$ \\
    & $\mu$ & $s_7$ & $[x_2 ~x_1 ~x_2]$ & $[\bar{v}_{62}~\bar{v}_{61}~\bar{v}_{62}]$ \\ \hline
    \multirow{4}{2em}{$s_7$} & $\lambda_b p_b$ & $s_7$ & $[x_0 ~x_1 ~ x_2]$ & $[\bar{v}_{70}~\bar{v}_{71}~\bar{v}_{72}]$\\
    & $\lambda_b \Bar{p}_b$ & $s_5$ & $[x_0 ~x_1 ~0]$ & $[\bar{v}_{70}~\bar{v}_{71}~0]$ \\
    & $\lambda^L_{b,h}$ & $s_7$ & $[x_0 ~x_0 ~x_2]$ & $[\bar{v}_{70}~\bar{v}_{70}~\bar{v}_{72}]$ \\
    & $\lambda^{b,h}$ & $s_6$ & $[x_0 ~x_1 ~x_0]$ & $[\bar{v}_{70}~\bar{v}_{71}~\bar{v}_{70}]$ \\
    & $\lambda_{e}$ & $s_7$ & $[x_0 ~x_1 ~x_0]$ & $[\bar{v}_{70}~\bar{v}_{71}~\bar{v}_{70}]$ \\
    & $\mu_0$ & $s_2$ & $[x_1 ~x_1 ~x_2]$ & $[\bar{v}_{71}~\bar{v}_{71}~\bar{v}_{72}]$ \\
    & $\mu$ & $s_7$ & $[x_2 ~x_1 ~x_2]$ & $[\bar{v}_{72}~\bar{v}_{71}~\bar{v}_{72}]$ \\ \hline
    \end{tabular}
    \caption{{State transitions of the CTMC and associated AoI jumps for Y/G-tasks.}}
    \label{table:transitions_yellow}
    \vspace{-2mm}
\end{table}

Now that the CTMC for the tasks with Y/G flags is completely characterized using Tables \ref{table:states_yellow} and \ref{table:transitions_yellow}, we next compute ${\Delta}^{(N)}_{b,i}(\bm{p},\bm{\mu_0}), b \in \{Y,G\}$ for $U_i$. Toward this end, let us define the quantity $\varrho_b:= \lambda_b + \lambda_{e} + \mu_0 + \mu + \lambda^h + \lambda^L_{b,h}$. Then, using \eqref{steady_state_prob}, the steady state vector $\Bar{\pi}_b$ for Y tasks satisfies \eqref{prob_sum} and the following set of equations,
\begin{align}\label{steady_state_prob_yellow}
    \varrho_b \Bar{\pi}_{b,1} \!\!& =\!\! (\lambda_b p_b + \mu_0 + \mu)\Bar{\pi}_{b,1} + \lambda_b p_b \Bar{\pi}_{b,2},  \\
    \varrho_b \Bar{\pi}_{b,2} \!\!& =\!\! (\lambda_b\Bar{p}_b \!+\! \mu_0 \!+\! \mu)\Bar{\pi}_{b,2} \!+\! \lambda_b \Bar{p}_b (\bar \pi_{b,1}\! + \!\Bar{\pi}_{b,4} ) \!+\! \mu_0(\Bar{\pi}_{b,5} \!+ \!\Bar{\pi}_{b,7}), \nonumber \\ 
    \varrho_b \Bar{\pi}_{b,3} \!\!& =\! \!(\lambda_b \!\!+\! \!\lambda^{b,h} \!\!+\!\! \lambda_{e} \!+\! \mu_0) \Bar{\pi}_{b,3} \!\!+\!\! \lambda^{b,h} (\Bar{\pi}_{b,1} \!+\! \Bar{\pi}_{b,2} \!+\! \Bar{\pi}_{b,4}) \!+\! \mu_0 \Bar{\pi}_{b,6}, \nonumber \\
    \varrho_b \Bar{\pi}_{b,4}\!\! & =\!\! (\lambda_b p_b + \mu_0 + \mu + \lambda_{e}) \Bar{\pi}_{b,4} + \lambda_{e} (\Bar{\pi}_{b,1} + \Bar{\pi}_{b,2}) + \mu \Bar{\pi}_{b,3}, \nonumber \\
    \varrho_b \bar \pi_{b,5} \!\!&= \!\!(\lambda_b + \lambda^L_{b,h} + \mu) \bar \pi_{b,5} + \lambda^L_{b,h} (\Bar{\pi}_{b,1} + \Bar{\pi}_{b,2}) + \lambda_b \bar p_b \Bar{\pi}_{b,7}, \nonumber \\
    \varrho_b \Bar{\pi}_{b,6}\!\! & =\!\! ( \lambda_b \!+\! \lambda^L_{b,h} \!+\! \lambda^{b,h} \!+\! \lambda_{e})\Bar{\pi}_{b,6} \!+\! \lambda^L_{b,h} \Bar{\pi}_{b,3} \!+\! \lambda^{b,h} (\Bar{\pi}_{b,5} \!+\! \Bar{\pi}_{b,7}), \nonumber \\
    \varrho_b \Bar{\pi}_{b,7} \!\!& =\! \! (\lambda_b p_b + \lambda_{e} \!+\! \lambda^L_{b,h} \!+\! \mu) \Bar{\pi}_{b,7} \!+\! \lambda^L_{b,h} \Bar{\pi}_{b,4} \!+\! \lambda_{e} \Bar{\pi}_{b,5} \!+\! \mu \Bar{\pi}_{b,6}.\!\!\nonumber
\end{align}
    
\begin{figure*}[h]
    \begin{align}\label{eqn_v_s_yellow}
        \varrho_b \Bar{v}_1  \!\!= &u_s \Bar{\pi}_{b,1} \!+\! \lambda_b{p}_b [\bar{v}_{10}~0~\bar{v}_{12}] + \mu_0 [\bar{v}_{11}~\bar{v}_{11}~\bar{v}_{11}] + \mu [\bar{v}_{12}~\bar{v}_{11}~\bar{v}_{12}] + \lambda_b p_b [\bar{v}_{20}~0~\bar{v}_{22}] \nonumber\\
        \varrho_b \Bar{v}_2  \!\!= & u_s \Bar{\pi}_{b,2} \!\!+\!\! \lambda_b \Bar{p}_b [\bar{v}_{20}~\bar{v}_{21}~0] \!\!+\!\! \mu [\bar{v}_{22}~\bar{v}_{22}~\bar{v}_{22}] \!\!+\!\! \lambda_b \Bar{p}_b [\bar{v}_{10}~\bar{v}_{11}~0] \!\!+\!\! \lambda_b \Bar{p}_b [\bar{v}_{40}~\bar{v}_{41}~0] \!\!+\!\! \mu_0 ([\bar{v}_{21}~\bar{v}_{21}~\bar{v}_{22}]\!\!+\!\![\bar{v}_{51}~\bar{v}_{51}~\bar{v}_{52}]  \!\!+\!\! [\bar{v}_{71}~\bar{v}_{71}~\bar{v}_{72}]) \nonumber \\
        \varrho_b \Bar{v}_3  \!\!= &u_s \Bar{\pi}_{b,3} + \lambda_b p_b [\bar{v}_{30}~0~\bar{v}_{32}] + \lambda_b \bar p_b [\bar{v}_{30}~\bar{v}_{31}~\bar{v}_{32}] + \lambda_{e} [\bar{v}_{30}~\bar{v}_{31}~\bar{v}_{32}] \!+ \!\mu_0 [\bar{v}_{31}~\bar{v}_{31}~\bar{v}_{31}] + \lambda^{b,h} [\bar{v}_{30}~\bar{v}_{31}~\bar{v}_{30}] \nonumber \\
        & + \lambda^{b,h} ([\bar{v}_{10}~\bar{v}_{11}~\bar{v}_{10}] + [\bar{v}_{20}~\bar{v}_{21}~\bar{v}_{20}] + [\bar{v}_{40}~\bar{v}_{41}~\bar{v}_{40}] ) + \!\mu_0 [\bar{v}_{61}~\bar{v}_{61}~\bar{v}_{62}]  \\
        \varrho_b \Bar{v}_4  \!\!=& u_s \Bar{\pi}_{b,4} \!\!+\!\! \lambda_b p_b [\bar{v}_{40}~0~\bar{v}_{42}] \!\!+\!\! \mu [\bar{v}_{42}~\bar{v}_{41}~\bar{v}_{42}] \!\!+ \!\!\mu_0 [\bar{v}_{41}~\bar{v}_{41}~\bar{v}_{41}] \!\!+\!\! \lambda_{e}( [\bar{v}_{40}~\bar{v}_{41}~\bar{v}_{40}]\!+\![\bar{v}_{10}~\bar{v}_{11}~\bar{v}_{10}]
         \!\! +\!\! [\bar{v}_{20}~\bar{v}_{21}~\bar{v}_{20}])\! \!+\!\! \mu [\bar{v}_{32}~\bar{v}_{31}~\bar{v}_{32}] \nonumber \\
        \varrho_b \Bar{v}_5  \!\!=& u_s \Bar{\pi}_{b,5} \!\!+\!\! \lambda_b p_b \bar v_5 \!\!+\!\! \lambda_b \bar p_b [\bar{v}_{50}~\bar{v}_{51}~0] \!\!+\!\! \lambda^L_{b,h} [\bar{v}_{50}~\bar{v}_{50}~\bar{v}_{52}] \!\!+\!\! \mu [\bar{v}_{52}~\bar{v}_{52}~\bar{v}_{52}] \!\!+ \!\! \lambda^L_{b,h}[\bar{v}_{10}~\bar{v}_{10}~\bar{v}_{12}]
         \!\! + \!\!\lambda^L_{b,h} [\bar{v}_{20}~\bar{v}_{20}~\bar{v}_{22}] \!\!+\!\! \lambda_b \bar p_b [\bar{v}_{70}~\bar{v}_{71}~0] \nonumber \\
        \varrho_b \Bar{v}_6  \!\!=& u_s \Bar{\pi}_{b,6} \!\!+ \!\!\lambda_b \bar v_6 \!\!+\!\! \lambda^L_{b,h} [\bar{v}_{60}~\bar{v}_{60}~\bar{v}_{62}]\!\!+\!\! \lambda^{b,h} [\bar{v}_{60}~\bar{v}_{61}~\bar{v}_{60}] \!\!+\! \!\lambda_{e} [\bar{v}_{60}~\bar{v}_{61}~\bar{v}_{62}] \!\!+\!\! \lambda^L_{b,h} [\bar{v}_{30}~\bar{v}_{30}~\bar{v}_{32}]
         \!  \!+\!\! \lambda^{b,h} ([\bar{v}_{50}~\bar{v}_{51}~\bar{v}_{50}] \!\!+\!\! [\bar{v}_{70}~\bar{v}_{71}~\bar{v}_{70}]) \nonumber \\
        \varrho_b \Bar{v}_7  \!\!=& u_s \Bar{\pi}_{b,7} \!\!+\!\! \lambda_b p_b \bar v_7\!\! + \!\!\lambda^L_{b,h} [\bar{v}_{70}~\bar{v}_{70}~\bar{v}_{72}]\!\!+\!\! \lambda_{e} [\bar{v}_{70}~\bar{v}_{71}~\bar{v}_{70}] \!\!+\!\! \mu [\bar{v}_{72}~\bar{v}_{71}~\bar{v}_{72}]\!\! + \!\!\lambda^L_{b,h} [\bar{v}_{40}~\bar{v}_{40}~\bar{v}_{42}]
          \!\!+\!\! \lambda_{e} [\bar{v}_{50}~\bar{v}_{51}~\bar{v}_{50}] \!\!+\!\! \mu [\bar{v}_{62}~\bar{v}_{61}~\bar{v}_{62}] \nonumber
    \end{align}
    \vspace{-4mm}
    \hrule
    % \vspace{-3mm}
\end{figure*}

Consequently, using \eqref{cond_prob_eqn}, we can write down the set of linear equations satisfied by the steady-state conditional distribution vector for Y tasks as in \eqref{eqn_v_s_yellow}. Then, we can state the following result which computes the expression of average AoI for the Y/G-tasks ($\Delta^{(N)}_{y,i}(\bm{p},\bm{\mu_0})$ and $\Delta^{(N)}_{g,i}(\bm{p},\bm{\mu_0})$) for UE $U_i$.

\begin{theorem}\label{thm:AoI_MEC_yellow}
    Suppose that the arrival rate of all tasks at UE $i$ is distributed as $\text{Poi}(\lambda_{a,i}), a \in \{R,Y,G\}$ 
    and the service rates of $L_i$ and ES are distributed as $\text{exp}(\mu_{0,i})$ and $\text{exp}(\mu)$, respectively. Then, the average AoI ${\Delta}^{(N)}_{b,i}(\bm{p},\bm{\mu_0})$ exists and is obtained by solving \eqref{steady_state_prob_yellow} and \eqref{eqn_v_s_yellow}.
\end{theorem}

Due to excessive lengths of the symbolic expressions of ${\Delta}^{(N)}_{y,i}(\bm{p},\bm{\mu_0})$ and ${\Delta}^{(N)}_{g,i}(\bm{p},\bm{\mu_0})$, we do not show them here. Later, we will provide an algorithm for computing the solution to Problem \ref{problem:N_user_game} without the need for closed-form expressions.

Theorems \ref{thm:AoI_MEC_red} and \ref{thm:AoI_MEC_yellow} can now be used to compute the weighted average AoI $\Delta_i^{(N)}(\bm{p},\bm{\mu_0})$ in Problem \ref{problem:N_user_game} for UE $i$, which then completes its definition. Next, we recall that the cost $J_i$ of UE $U_i$ depends on the offloading policies of the other UEs, which couples its own optimization problem with that of the others. This would, in general, require UEs to share their policy information amongst themselves, which can cause significant overhead and may not even be possible in a large population spatially distributed setting. Thus, to address this challenge where each UE wants to make completely decentralized decisions, we employ the MFG framework to compute an approximate Nash equilibrium (NE) solution.

\section{The MEC Mean-Field Game}
Our aim now is to compute NE policies $(p_i^*,\mu_{0,i}^*)$ for each UE $U_i$ that satisfy the following set of inequalities \cite{bacsar1998dynamic}:
\begin{align*}
    & J_i(p_i^*,\mu_{0,i}^*,p_{-i}^*,\mu_{0,-i}^*) \leq J_i(p_i,\mu_{0,i},p_{-i}^*,\mu_{0,-i}^*),~\forall i \in [N].
\end{align*}

Briefly, according to the above set of inequalities, any \textit{rational} UE who tries to deviate from the NE policy $(\bm{p}^*,\bm{\mu_0}^*)$ incurs a higher cost, and thus, it is in the best interest of each UE to follow it. However, finding such policies requires that each UE has access to the policies of the other UEs, which as noted (a) can be difficult to obtain in a high user population setting, and (b) can incur significant communication overhead. Thus, to alleviate these issues, we design \textit{decentralized} NE policies for the UEs where each UE utilizes only its own \textit{local information} and statistical information of the underlying MEC system. In this regard, we leverage the framework of MFG \cite{al2015joint}. Motivated by statistical physics, a MFG approximates the complex interactions in an interacting UE system with an \textit{average effective field} (also referred to as the mean-field (MF)) using a countably infinite population ($N = \infty$). Subsequently, each UE now reacts to the MF generated by the entire UE population (rather than each other individual UE) in a manner such that its own behavior is consistent with that of the other UEs. Then, to derive equilibrium solutions, it suffices to consider the perspective of one \textit{generic} UE, which represents each type of the population.
We now set up the MEC MFG.

Consider a generic UE of type $\phi \in \Phi$. The tasks arriving at the generic device are distributed as $\text{Poi}(\lambda_{\phi,a})$ with $a \in \{R,Y,G\}$. Each of these are either executed using the local processor or offloaded to the ES by employing a mean $p_{\phi,a}$, i.i.d.~Bernoulli distributed. This splits the incoming arrival process into two independent Poisson processes with respective means $\lambda_{\phi,a} p_{\phi,a}$ and $\lambda_{\phi,a} \Bar{p}_{\phi,a}$. Further, the service times of the generic UE's local processor is distributed as exponential with parameter $\mu_{0,\phi} \leq f_{\phi,max}$.

Next, let us define $p_\phi: = (p_{\phi,r},p_{\phi,y},p_{\phi,g})$, $\lambda^a:= \sum_{j \in [N]}\lambda_{a,j} \bar p_{a,j}, a \in \{R,Y,G\}$ and the quantities:
\begin{align*}
    \rho_{a}^{(N)} & := \frac{\lambda^a}{(N-1) \mu} ~\text{ and }~
\rho_a := \lim_{N \rightarrow \infty}\rho_a^{(N)},
\end{align*}
where $\rho_a^{(N)}$ serves to denote the mean loading on the ES due to an incoming task with flag $a$ and $\rho_a$ denotes the MF approximation (or the aggregate loading due to an infinite number of UEs) of $\rho^{(N)}_a$ as discussed earlier. Then, we have that for a high UE MEC system, the arrival rates $\lambda_{-a}$ and $\lambda^{a}$ can be approximated as:
\begin{align*}
    \lambda_{-a} & = (N-1) \mu *\frac{\lambda_{-a}}{ (N-1) \mu}\approx (N-1) \mu \rho_a,a \in \{R,Y,G\} \\
    \lambda^a & = N \mu *\frac{\lambda^a}{ N \mu}\approx N \mu \rho_a,~~~a \in \{R,Y\}.
\end{align*}

With the above resubstitutions, we make an important observation that $\lambda_{-a}$ and $\lambda^a$ are no longer dependent on the offloading policies of UEs of other types (as was the case in the definitions proposed in Subsections \ref{subsec:red} and \ref{subsec:yellow}), but only on the aggregate loading at the ES. This serves to decouple the otherwise coupled optimization problems of the UEs (stated in Problem \ref{problem:N_user_game}) which can now be solved for \textit{a priori} given loading $\rho_a$, for all $a$, in addition to an extra consistency condition as we will see next.

Define $\rho:= (\rho_r,\rho_y,\rho_g)$. Consequently, the average AoI of a generic UE of type $\phi$ can be computed as:
\begin{align}\label{MF_AoI}
    & {\Delta}_{\phi,\rho}(p_\phi,\mu_{0,\phi}) \!:=\! \Delta^{(N)}_{\phi_i}(\bm{p},\bm{\mu_0};\lambda_{-a},\lambda^a) \mid_{\substack{\lambda_{-a} = (N-1) \rho_a \mu \\ \lambda^a = N \rho_a \mu~~~}},
\end{align}
for all $\phi \in \Phi$, where the notation $x(z)\mid_{z = b}$ denotes the value of $x$ when $b$ is substituted for the argument $z$. Then, we can formally state a generic UE's optimization problem as below.

\begin{problem}[Generic UE optimization problem]\label{problem:generic_user}
    \begin{align*}
        \min_{(p_\phi,\mu_{0,\phi}) \in [0,1]^3 \times \mathbb{R}} &  J_{\rho}(p_\phi,\mu_{0,\phi}) :=P_{\ell,\phi} + V {\Delta}_{\phi,\rho}(p_\phi,\mu_{0,\phi}) \\
        & \hspace{-9mm}\mbox{s.t.}~~~ \mu_{0,\phi} \leq f_{\phi,max}.
        \end{align*} 
\end{problem}

Hence, the MEC MFG is defined over two operators, namely, the optimality and the consistency operators as:
\begin{enumerate}
    \item \textit{Optimality:} $(\hat{p}_\phi,\hat{\mu}_{0,\phi}) = \Psi_1(\rho) := \text{argmin} ~J_\rho(p_\phi,\mu_{0,\phi})$     subject to the constraint in Problem \ref{problem:generic_user}.
    \item \textit{Consistency:} $\forall a \in \{R,Y,G\}$, we have
    \begin{align*}
        (\hat \rho)_a & = \Psi_2(\hat{p}_\phi,\hat{\mu}_{0,\phi}) := \Big(\frac{1}{\mu}\mathbb{E}_{\phi}[\lambda_{\phi,a} \bar p_{\phi,a}]\Big)_{a}.
    \end{align*}
\end{enumerate}
The mean-field equilibrium (MFE) which constitutes the tuple of equilibrium policies $(p_{\phi,\text{MFE}},\mu_{0,\phi,\text{MFE}})_{\forall \phi}$ and the equilibrium mean-field ($\rho_{\text{MFE}}$), is given by the fixed point of the composite map of $\Psi_1$ and $\Psi_2$. 

Next, we note that the non-convex nature of the cost function $J_\rho$ entails that the resulting MFG is a non-convex game. For this, we provide a low complexity algorithm (Algorithm \ref{alg:MFE}) to compute (local) MFEs for the MEC game. Using this, we will carry out numerical evaluation of the proposed MFG approach to assess the merits of the approach.

\begin{algorithm}[h!]
	\caption{Fixed point iteration for computing MFE policy}
    \label{alg:MFE}
	\begin{algorithmic}[1]
        \STATE {\textbf{Input:} $V,\eta_\phi,\lambda_\phi,\mu,w_r,w_y,w_g, ~\forall \phi$ \hfill\# System parameters}
        \STATE {\textbf{Input:} $\varepsilon_1, \varepsilon_2$ \hfill \# tolerance parameters}
        \STATE {\textbf{Input:} $\gamma_1,\gamma_2$ \hfill \# Iteration step sizes}
		\STATE {Initialize: $\hat{\rho}^{(0)}$, $\sigma_\phi^{(0)}:= (p_\phi^{(0)},\mu_{0,\phi}^{(0)}), \forall \phi$}
        \STATE{$k \gets 1$}
        \WHILE{$|\hat{\rho}^{(m)} - \hat{\rho}^{(m-1)}| > \varepsilon_1$}
        \FOR{$\phi \in \Phi$}
        \FOR{{$i \in \{1,2,3,4\}$}} \hfill \# Coordinate-wise descent
        \STATE{$\ell \gets 1$}
        \WHILE{$|\hat{\sigma}_\phi^{(m')}(i) - \hat{\sigma}_\phi^{(m'-1)}(i)| > \varepsilon_2$}
        \STATE $\hat{\sigma}_\phi^{(\ell)}\!(i)\!\leftarrow \!\hat{\sigma}_\phi^{(\ell-1)}\!(i) \!-\! \gamma_2\nabla \! J_{\hat{\rho}^{(k-1)}}(\sigma_\phi^{(k-1)})$
        \STATE{$\ell \gets \ell + 1$}
        \ENDWHILE
        \STATE $ \sigma_\phi^{(k)}(i) = \hat \sigma_\phi^{(m')}(i)$
        \ENDFOR
        \ENDFOR
        \STATE $\hat{\rho}^{(k)} \leftarrow (1-\gamma_1)\hat{\rho}^{(k-1)} + \gamma_1 \Big(\mathbb{E}_\phi \Big[\frac{\lambda_{\phi,a}(1-{p}^{(k)}_{\phi,a})}{\mu}\Big]\Big)_{\forall a}$
        \STATE{$k \gets k+1$}
        \ENDWHILE
        \STATE \textbf{Output: } $\hat{\rho}^{(m)}, \sigma_\phi^{(m)}, ~\forall \phi.$
	\end{algorithmic}
    % \vspace{-2mm}
\end{algorithm}

Briefly, Algorithm \ref{alg:MFE} computes a fixed point of the composite operator $\Psi_2 \circ \Psi_1$. Thus, given a value of $\rho$, we solve a generic user's optimization problem defined in Problem \ref{problem:generic_user} using block coordinate gradient descent method \cite{wright2015coordinate} (lines 8-12). Subsequently, we update the mean-field via using the obtained optimal policy (line 15). The algorithm iterates until convergence and the MFE is given by the last iterate (line 18).

\section{Performance Evaluation}
\begin{figure}[t]
    \centering    \includegraphics[width=\columnwidth]{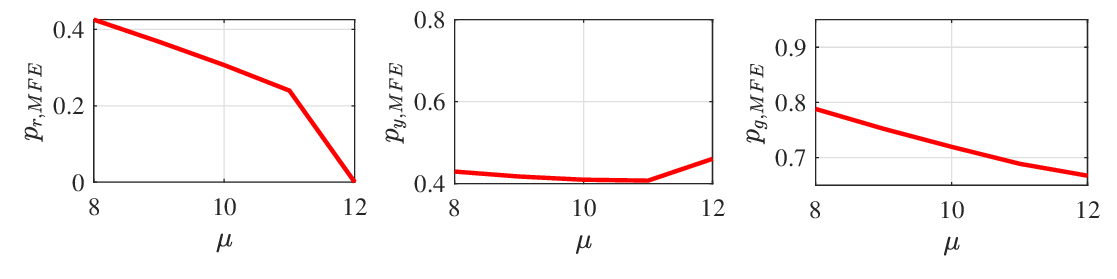}
        % \vspace{-8mm}
    \caption{{The variation of equilibrium probability of local processing for each of the red, yellow, and green tasks versus the ES parameter $\mu$.}}
    \label{fig:load_vs_prob}
    \vspace{-4mm}
\end{figure}

Now, we assess the merits of the proposed approach through numerical evaluation. Due to page limitations, we restrict ourselves to two case studies only. In the first one, we study the effect of increase in the ES service rate on the probability of local service in Fig. \ref{fig:load_vs_prob}. System parameters are taken to be: $\lambda_{r} = 1, \lambda_y = 3, \lambda_g = 6, N = 10, V = 10, w_r = 20, w_y = 5, w_g = 2$ and $f_{max} = 2$. Further, the initial conditions are set to $\mu_0 = 0.7$, $p_r = 0.6,$ $ p_y = 0.5$, and $p_g = 0.6$. From the figure, we observe that, in general, the probability of serving locally decreases and hence, that of offloading (which is $1-p_a$) increases with the increase in ES parameter $\mu$ (for Y-tasks, it peaks slightly). The variation is more prominent for R-tasks due to server saturation and priority in service. Meanwhile, the Y/G tasks are observed to be offloaded more cautiously (to prevent increased preemptions at ES due to R-tasks) and show minor variations when plotted on the same y-axes scale.
% and as a resultHowever, a noteworthy observation is the level of increase for R/Y/G tasks. Due to higher weights on the average AoI for R/Y, these are more cautiously offloaded compared to the G tasks in which we observe much higher difference in offloading as $\mu$ increases. The R-tasks are however, offloaded more and more compared to the yellow ones since they have priority in service as well as a much lower incoming rate compared to the yellow ones. 

\begin{figure}[t]
    \centering    \includegraphics[width=\columnwidth]{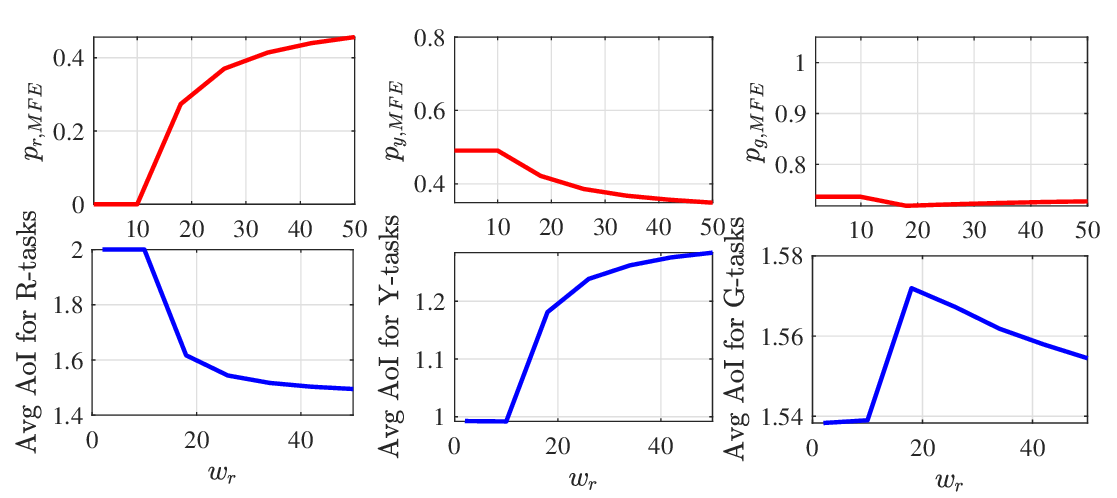}
    \caption{{The equilibrium probabilities of local processing for each of the red, yellow and green tasks as a function of the weight $w_r$ for the R-tasks.}}
    \label{fig:load_vs_wr}
    \vspace{-4mm}
\end{figure}

Next, for the same list of system parameters as above and $\mu = 10$, we study the effect of increasing the weighting $w_r$ for the AoI of the R-tasks in Fig. \ref{fig:load_vs_wr}. The leftmost subfigure in the same shows that the offloading probability decreases with increasing $w_r$ due to a higher weight on freshness of information. Keeping the same scale on y-axes in the first row subfigures, we observe that the offloading probability of Y-tasks increases slightly. This is because of the decrease in R-tasks offloading which gives more opportunity for completion of the Y-tasks. There is no significant change in the offloadings of the G-tasks, although a careful observation shows a valley point around $w_r = 20$, where the cost of power consumption balances the overall AoI cost, after which, if gets expensive to offload, and thus users tend to serve locally. Finally, we would also like to note that the obtained results can be highly dependent on the parameters of the problem, and there can be varying trends for varying system parameters. 

\section{Conclusion}
We have designed decentralized offloading policies for $N$ users in a MEC system with one ES assisting them for processing computation intensive tasks. There are three streams of tasks incoming at each UE, namely red (of highest urgency), yellow (of moderate urgency), and green (of lowest urgency). We first formulated the computation offloading problem as an $N$-user noncooperative game problem and consequently utilized the MFG framework to compute completely decentralized policies for each user to balance between power consumption due to local processor usage and average AoI incurred as a result of resource sharing at the ES. In the process, we also computed closed-form expressions for the average AoIs incurred by different types of tasks using techniques from SHS theory. Using simulations, we corroborated the theoretical results to observe that using higher weighting on the AoI portion of the cost may lead to more local processor utilization, while increasing ES capacity allows UEs to push more computations towards the edge, aligned with intuition.

\bibliographystyle{IEEEtran}
\bibliography{references}

\end{document}